\documentclass[11pt]{article}

\usepackage[utf8]{inputenc}
\usepackage[T1]{fontenc}
\usepackage{amsmath,amssymb,amsfonts}
\usepackage{bm}
\usepackage{geometry}
\usepackage{hyperref}
\usepackage{graphicx}
\usepackage{authblk}
\usepackage{cite}
\usepackage{microtype}

\newcommand{\dd}{\mathrm{d}}
\newcommand{\ee}{\mathrm{e}}
\newcommand{\ii}{\mathrm{i}}

\newcommand{\cL}{\mathcal{L}}
\newcommand{\cZ}{\mathcal{Z}}
\newcommand{\cO}{\mathcal{O}}
\newcommand{\cS}{\mathcal{S}}
\newcommand{\cH}{\mathcal{H}}
\newcommand{\avg}[1]{\left\langle #1 \right\rangle}
\newcommand{\ket}[1]{\left|#1\right\rangle}
\newcommand{\bra}[1]{\left\langle #1 \right|}

\newcommand{\cG}{\mathcal{G}}

\hypersetup{
colorlinks=true,
linkcolor=blue,
citecolor=blue,
urlcolor=blue
}

\title{Comminution as a Non-Hermitian Quantum Field Theory:\
Log-Size Jump Generators, Branching Embeddings, and the Airy Solvable Sector}

\author[1]{Juan J. Segura, Universidad Andr\'es Bello\\juan.segura.f@unab.cl}
\date{November 19, 2025}

\begin{document}
\maketitle

\begin{abstract}
Pure-breakage population balance equations (PBEs) give the standard deterministic description of fragmentation and comminution. They predict mean particle size distributions, but they do not determine fluctuations, size-size correlations, or universality under coarse-graining. We develop a field-theoretic framework anchored in the PBE kernel inputs (selection rate and daughter distribution) and compatible with the stochastic Doi-Peliti approach.

From homogeneous kernels we derive an exact Markov jump generator in log-size for a mass-weighted (tagged-mass) distribution, with a jump law that is a probability density fixed by the daughter distribution. The generator is generically non-self-adjoint, admits a Lindblad embedding, and has a second-quantized extension. The deterministic PBE appears as the one-body sector, while multi-point correlators encode finite-population fluctuations. We also give a binary-fragmentation embedding whose mean-field limit reproduces the PBE but whose higher correlators capture multiplicative cascade noise. For a linear Airy-type kernel, long-wavelength coarse-graining yields an effective Airy operator as a solvable quadratic sector about a stationary profile, producing explicit mode-sum formulas for equal-time connected two-point correlations. Overall, the framework separates what is fixed by kernel data from what requires additional stochastic modeling and links comminution kernels to universality classes.
\end{abstract}

\section{Introduction}

Fragmentation and comminution are central to particle processing, geophysics, aerosols, and
materials science. A large fraction of modelling practice rests on deterministic pure-breakage
population balance equations (PBEs), which evolve a size-resolved density under a breakage rate
(selection function) and a daughter kernel describing the distribution of fragments produced by a
break event. For many homogeneous (self-similar) kernels, the resulting particle size distributions
(PSDs) exhibit scaling forms and convergence to asymptotic profiles, with a substantial
mathematical theory underpinning existence, uniqueness, and rates of convergence in appropriate
function spaces~\cite{ChengRedner1988,ChengRedner1990,Escobedo2005,Caceres2010,Caceres2011,Biedrzycka2019,BertoinBook}.

At the same time, three broad questions remain comparatively underdeveloped in comminution-oriented
PBE modelling:

\begin{enumerate}
\item \textbf{Fluctuations and correlations.} PBEs determine the \emph{mean} evolution of a density,
but do not specify the stochastic structure that produces fluctuations, covariance matrices, or
higher cumulants of PSD measurements.

\item \textbf{Non-Hermitian operator structure and solvable sectors.} Several kernel families admit
useful transformations to drift--diffusion operators and, after similarity transforms, to
non-self-adjoint Schr"odinger-type operators whose eigenmodes organise PSD families. This
spectral perspective becomes more powerful when it extends beyond the one-body level.

\item \textbf{Universality under coarse-graining.} Comminution is a paradigmatic non-equilibrium,
multiplicative cascade. Field-theoretic tools developed for branching processes, absorbing-state
transitions, and growth--fragmentation models~\cite{Doi1976,Peliti1985,CardyTauber1998,Hinrichsen2000,Odor2004,Tauber2014Book}
suggest that universality arguments could be brought to bear, but a transparent bridge from
engineering kernels to field-theory actions is often missing.
\end{enumerate}

In earlier work, a class of homogeneous pure-breakage PBEs was mapped---under explicit
coarse-graining assumptions---to a non-Hermitian quantum-mechanical (NHQM) description in log-size,
where a PSD snapshot is read as a squared wavefunction profile $|\psi(\xi)|^2$ of an effective
non-Hermitian operator. The present article advances that program in two directions that are
necessary for a publishable, foundational framework:

\begin{itemize}
\item We make \emph{explicit} the stochastic modelling step that is usually implicit when one
discusses fluctuations: from a deterministic PBE to a microscopic Markov process consistent with
the PBE at the one-body level.

\item We then show how that Markov generator admits both (i) a Lindblad embedding (ensuring
positivity and conservation laws at the operator level) and (ii) a second-quantized extension to a
bosonic NHQFT whose correlators encode PSD fluctuations.
\end{itemize}

\subsection*{Two complementary stochastic embeddings}

A key gap in many ``field theory from PBE'' discussions is that \emph{the deterministic PBE does not
uniquely determine the noise}. Distinct microscopic fragmentation mechanisms can share the same
one-body equation but differ in higher statistics. We therefore treat two embeddings in parallel:

\paragraph{(A) Tagged-mass (linear) embedding.}
One samples an infinitesimal mass element (a ``tagged mass quantum'') and tracks the log-size of
the fragment containing it. A breakage event moves the tag to a smaller daughter with a
\emph{mass-weighted} transition probability. This yields a \emph{linear} Markov jump equation for a
normalised density $p(\xi,t)$. The associated NHQFT is Gaussian (free) apart from nonlocality, and
serves as a controlled baseline for fluctuation calculations.

\paragraph{(B) Particle-number (branching) embedding.}
One tracks the particles themselves: a parent breaks into a random set of daughters (for
concreteness, binary splitting), producing a genuine branching process. The mean field reproduces
the classical PBE, while higher correlators encode cascade noise and intermittency. The associated
NHQFT is interacting (branching vertex) and naturally comparable to established branching
field theories.

Both embeddings are useful. Embedding~(A) is the cleanest continuation of the NHQM picture and is
fully determined by standard kernel data. Embedding~(B) addresses the physically intuitive fact
that fragmentation increases particle number and supplies the correct nonlinearity structure for
universality discussions.

\subsection*{Outline}

Section~\ref{sec:pbe-to-log} fixes conventions and derives the exact log-size generator for
homogeneous kernels, with particular care devoted to conservation and normalisation.
Section~\ref{sec:lindblad-onebody} presents the Lindblad embedding and the non-Hermitian operator
viewpoint. Section~\ref{sec:nhqft} constructs the second-quantized NHQFT and its Doi--Peliti
representation, first for the tagged-mass theory and then for a binary-branching extension.
Section~\ref{sec:airy} specialises to the Airy-type kernel and develops a solvable quadratic sector
and a Gaussian two-point function. Section~\ref{sec:universality} discusses universality and RG
outlook with explicit caveats. Section~\ref{sec:conclusions} summarises results and open problems.
Technical derivations are collected in appendices.

\begin{figure}[t]
\centering
\caption{Schematic mapping developed in this article. Starting from a homogeneous pure-breakage
PBE, one may: (i) form a log-size jump generator for a mass-weighted (tagged-mass) density,
which admits a Lindblad embedding and a free (Gaussian) NHQFT; and/or (ii) specify a microscopic
branching rule (e.g.\ binary splitting) whose mean-field equation reproduces the PBE, yielding an
interacting branching NHQFT. In both cases, coarse-graining can produce solvable quadratic
sectors (e.g.\ Airy) that organise PSD profiles and fluctuation spectra.}
\label{fig:mapping}
\end{figure}

\section{Homogeneous PBEs and an Exact Log-Size Jump Generator}
\label{sec:pbe-to-log}

\subsection{Deterministic fragmentation equation and homogeneous kernels}

Let $f(x,t)$ denote the \emph{number density} of particles of size $x>0$.
A standard pure fragmentation (pure-breakage) equation reads
\begin{equation}
\partial_t f(x,t)= -S(x),f(x,t)
\int_x^\infty b(x,y),S(y),f(y,t),\dd y.
\label{eq:pbe-number}
\end{equation}
Here $S(x)\ge 0$ is the breakage (selection) rate and $b(x,y)\ge 0$ is the \emph{daughter number
kernel}: for a parent of size $y$, the expected number of fragments produced in $(x,x+\dd x)$ is
$b(x,y),\dd x$.

Mass conservation (with ``mass'' proportional to $x$) is enforced by
\begin{equation}
\int_0^y x,b(x,y),\dd x = y.
\label{eq:mass-cons-number}
\end{equation}
Consequently, the total mass
\begin{equation}
M := \int_0^\infty x,f(x,t),\dd x
\label{eq:total-mass}
\end{equation}
is conserved for solutions with sufficient integrability; the total number
$N(t)=\int_0^\infty f(x,t)\dd x$ typically increases.

We focus on the homogeneous kernel family
\begin{equation}
S(x)=k,x^\alpha,
\qquad
b(x,y)=\frac{1}{y},B!\left(\frac{x}{y}\right),
\qquad 0<x<y,
\label{eq:hom-kernel}
\end{equation}
with $k>0$, $\alpha\in\mathbb{R}$, and a dimensionless daughter density $B(z)$ on $z\in(0,1)$.
Condition~\eqref{eq:mass-cons-number} becomes
\begin{equation}
\int_0^1 z,B(z),\dd z = 1.
\label{eq:B-normalisation}
\end{equation}

\subsection{Mass-weighted log-size density and exact master equation}

Introduce the log-size coordinate
\begin{equation}
\xi = \ln(x/x_0), \qquad x=x_0,\ee^\xi,
\label{eq:log-size}
\end{equation}
and define the \emph{mass-weighted} (tagged-mass) log-size density
\begin{equation}
p(\xi,t)
:= \frac{x^2 f(x,t)}{M}
= \frac{x_0^2 \ee^{2\xi} f(x_0\ee^\xi,t)}{M},
\qquad \int_{-\infty}^{\infty} p(\xi,t),\dd \xi = 1.
\label{eq:p-log}
\end{equation}
This choice is not cosmetic: $p(\xi,t)\dd\xi$ is the fraction of \emph{total mass} carried by
particles whose log-size lies in $(\xi,\xi+\dd\xi)$. It also coincides with the law of a tagged
mass element under a Markovian fragmentation model (Appendix~\ref{app:tagged-mass}).

For homogeneous kernels~\eqref{eq:hom-kernel}, $p(\xi,t)$ obeys an \emph{exact} log-size jump equation
\begin{equation}
\partial_t p(\xi,t)
= -\lambda(\xi),p(\xi,t)
\int_0^{\infty}\lambda(\xi+u),K(u),p(\xi+u,t),\dd u,
\label{eq:log-master}
\end{equation}
with the (log-size) breakage rate
\begin{equation}
\lambda(\xi) = k,x_0^\alpha,\ee^{\alpha\xi}
\label{eq:lambda}
\end{equation}
and a \emph{normalised} log-jump kernel
\begin{equation}
K(u) = \ee^{-2u},B(\ee^{-u}), \qquad u\ge 0,
\qquad \int_0^\infty K(u),\dd u = 1.
\label{eq:Kdef}
\end{equation}
The normalisation of $K$ follows directly from~\eqref{eq:B-normalisation} by the change of
variables $z=\ee^{-u}$:
$\int_0^\infty K(u)\dd u = \int_0^1 z,B(z)\dd z = 1$.

\paragraph{Interpretation.}
Equation~\eqref{eq:log-master} is a forward Kolmogorov equation for a pure jump process on the line:
between events, $\xi$ is constant; at rate $\lambda(\xi)$ it jumps to $\xi-u$, where $u\ge 0$ is
distributed with density $K(u)$. In particular, \emph{the deterministic PBE data uniquely determine
the tagged-mass jump law} $K$; no additional noise modelling is required at the one-body level.

\subsection{Small-jump expansion and the drift--diffusion generator}

To connect with NHQM-type differential operators, one may expand the gain term of
\eqref{eq:log-master} in moments of $u$. Define
\begin{equation}
m_n := \int_0^\infty u^n,K(u),\dd u \qquad (n=1,2,\dots).
\label{eq:Kn-moments}
\end{equation}
Assuming $K(u)$ is concentrated near $u=0$ and that $\lambda(\xi)p(\xi,t)$ varies smoothly on the
jump scale, a Kramers--Moyal expansion yields
\begin{equation}
\partial_t p(\xi,t)= m_1,\partial_\xi!\bigl(\lambda(\xi)p(\xi,t)\bigr)
\frac{m_2}{2},\partial_\xi^2!\bigl(\lambda(\xi)p(\xi,t)\bigr)
\cO(m_3\partial_\xi^3).
\label{eq:KM}
\end{equation}
Truncating at second order gives a Fokker--Planck form
\begin{equation}
\partial_t p = -\partial_\xi!\bigl(v(\xi),p\bigr) + \partial_\xi^2!\bigl(D(\xi),p\bigr),
\qquad
v(\xi) = -m_1,\lambda(\xi),
\qquad
D(\xi) = \frac{m_2}{2},\lambda(\xi).
\label{eq:FP}
\end{equation}
This is a controlled long-wavelength limit when $m_1,m_2$ are finite and higher cumulants are
small on the scales probed. The corresponding operator is non-self-adjoint unless $v$ and $D$ are
related by detailed-balance conditions (not expected here), motivating the ``non-Hermitian''
language.

\paragraph{From Fokker--Planck to Schr"odinger form.}
A standard similarity transform converts~\eqref{eq:FP} to an imaginary-time Schr"odinger-type
equation for an amplitude $\psi(\xi,t)$ (Appendix~\ref{app:schr-map}). Under analytic continuation
one obtains a non-Hermitian Schr"odinger equation of the general form
\begin{equation}
\ii,\hbar,\partial_t \psi(\xi,t)
\left[-\frac{\hbar^2}{2\mu}\partial_\xi^2 + V(\xi) - \ii,\Gamma(\xi)\right]\psi(\xi,t),
\label{eq:schrodinger-general}
\end{equation}
with $p(\xi,t)=\ee^{-\Phi(\xi)}|\psi(\xi,t)|^2$ for a gauge $\Phi(\xi)$ fixed by $(v,D)$.
The present article treats~\eqref{eq:log-master} as the \emph{fundamental} object and views
\eqref{eq:schrodinger-general} as an effective, coarse-grained representation.
\section{Lindblad Embedding of the Log-Size Generator}
\label{sec:lindblad-onebody}

\subsection{Operator form and non-Hermitian generator}

Introduce a log-size Hilbert space spanned by kets $\ket{\xi}$ with resolution of identity
$\int \dd\xi,\ket{\xi}\bra{\xi} = \mathbb{I}$.
Define a diagonal density operator
\begin{equation}
\rho(t) = \int_{-\infty}^{\infty} p(\xi,t),\ket{\xi}\bra{\xi},\dd\xi.
\label{eq:rho-diag}
\end{equation}
The jump generator in~\eqref{eq:log-master} can be written as a linear map $\partial_t\rho=\cG[\rho]$,
with $\cG$ non-self-adjoint in general. Writing the generator as ``non-Hermitian Hamiltonian'' is
common in both stochastic processes and non-equilibrium field theory: it simply reflects that the
evolution operator is not unitary and its spectrum need not be real.

\subsection{Completely positive Lindblad embedding}
A key structural advantage of using a Lindblad form is that complete positivity is built in. For
each jump size $u\ge 0$ define a single-particle jump operator
\begin{equation}
L(u) = \int_{-\infty}^{\infty}
\sqrt{\lambda(\xi),K(u)};\ket{\xi-u}\bra{\xi},\dd\xi,
\qquad u\ge 0.
\label{eq:L-single}
\end{equation}
Then the \emph{pure-jump} evolution~\eqref{eq:log-master} is reproduced by the Lindblad master equation
\begin{equation}
\partial_t\rho
\int_0^\infty
\left(
L(u)\rho L^\dagger(u) - \frac{1}{2}{L^\dagger(u)L(u),\rho}
\right)\dd u,
\label{eq:Lindblad-single}
\end{equation}
provided $\rho$ is initialised diagonal in the $\ket{\xi}$ basis. In that case, the diagonal
elements of~\eqref{eq:Lindblad-single} are exactly~\eqref{eq:log-master}.

\paragraph{Remark (where ``non-Hermiticity'' lives).}
Equation~\eqref{eq:Lindblad-single} has no Hamiltonian part; all non-unitarity is encoded in the
jump terms. If one introduces additional coarse-grained drift/diffusion as a \emph{coherent} term,
one may add a Hermitian Hamiltonian $\cH$ to the commutator part. The effective non-Hermitian
operator of NHQM then appears as the standard no-jump Hamiltonian
$H_{\rm eff}=H-\frac{\ii\hbar}{2}\int_0^\infty L^\dagger(u)L(u)\dd u$, plus possible extra sinks.
We will keep the stochastic-generator viewpoint explicit to avoid conceptual confusion: the core
object is the Markov generator; the NHQM form is a useful coarse-grained representation.

\section{Second Quantisation: A Non-Hermitian Quantum Field Theory on Log-Size}
\label{sec:nhqft}

\subsection{Bosonic fields and observables}

Promote log-size to a continuous label of bosonic creation and annihilation fields
$a^\dagger(\xi)$ and $a(\xi)$ satisfying
\begin{equation}
[a(\xi),a^\dagger(\eta)] = \delta(\xi-\eta),
\qquad
[a(\xi),a(\eta)]=[a^\dagger(\xi),a^\dagger(\eta)]=0.
\label{eq:CCR}
\end{equation}
Define the number density operator in log-size space
\begin{equation}
\hat n(\xi) = a^\dagger(\xi)a(\xi),
\label{eq:n-operator}
\end{equation}
and let $\avg{\cdot}$ denote expectation with respect to a Fock-space density operator $\rho_F(t)$.

In the tagged-mass embedding, the total number of quanta $N=\int\hat n(\xi)\dd\xi$ is a modelling
parameter (e.g.\ number of tagged mass packets used to represent the continuum mass density). The
one-body PSD is obtained as
\begin{equation}
p(\xi,t) = \frac{1}{N},\avg{\hat n(\xi,t)},
\qquad \int p(\xi,t)\dd\xi=1.
\label{eq:p-from-n}
\end{equation}
Finite-$N$ fluctuations and correlations appear in connected correlators such as
\begin{equation}
G_c(\xi_1,\xi_2;t)
\avg{\hat n(\xi_1,t)\hat n(\xi_2,t)}
\avg{\hat n(\xi_1,t)},\avg{\hat n(\xi_2,t)}.
\label{eq:Gc}
\end{equation}

\subsection{Tagged-mass NHQFT: second-quantised jump Lindblad}

The natural Fock-space lift of~\eqref{eq:L-single} is the continuum family
\begin{equation}
\hat L(u)
\int_{-\infty}^{\infty}
\sqrt{\lambda(\xi),K(u)};
a^\dagger(\xi-u),a(\xi),\dd\xi,
\qquad u\ge 0.
\label{eq:L-Fock-hopping}
\end{equation}
We define the tagged-mass NHQFT by the Lindblad equation
\begin{equation}
\partial_t \rho_F
\int_0^\infty
\left(
\hat L(u)\rho_F \hat L^\dagger(u)
-\frac{1}{2}{\hat L^\dagger(u)\hat L(u),\rho_F}
\right)\dd u.
\label{eq:Lindblad-Fock}
\end{equation}
A direct calculation using~\eqref{eq:CCR} gives the exact one-body equation
\begin{equation}
\partial_t \avg{\hat n(\xi)}-\lambda(\xi)\avg{\hat n(\xi)}+\int_0^\infty \lambda(\xi+u)K(u)\avg{\hat n(\xi+u)},\dd u,
\label{eq:n-evol-hopping}
\end{equation}
so that~\eqref{eq:p-from-n} reproduces~\eqref{eq:log-master}.

\paragraph{What is new here (beyond deterministic PBEs).}
Equation~\eqref{eq:n-evol-hopping} is the closed equation for the \emph{mean} density. However, the
same NHQFT fixes a full hierarchy for equal-time and multi-time correlators, which capture the
finite-$N$ fluctuation structure associated with the tagged-mass jump process. This information is
absent from the deterministic PBE unless a stochastic embedding is specified.

\subsection{Doi--Peliti representation of the tagged-mass theory}

The tagged-mass Lindblad generator is equivalent (after the standard Doi shift) to a Doi--Peliti
functional integral. Introducing coherent-state fields $\varphi(\xi,t)$ and response fields
$\tilde\varphi(\xi,t)$, one obtains
\begin{equation}
\cZ = \int \mathcal{D}\varphi,\mathcal{D}\tilde\varphi;\exp!\bigl[-\cS_{\rm hop}[\varphi,\tilde\varphi]\bigr],
\end{equation}
with action
\begin{equation}
\cS_{\rm hop}
\int \dd t \int \dd\xi;
\tilde\varphi(\xi,t),\partial_t \varphi(\xi,t)
+
\int \dd t \int \dd\xi \int_0^\infty \dd u;
\lambda(\xi)K(u),
\Bigl[\tilde\varphi(\xi,t)-\tilde\varphi(\xi-u,t)\Bigr]\varphi(\xi,t).
\label{eq:DP-hop}
\end{equation}
The mean-field equation $\delta \cS/\delta\tilde\varphi=0$ is exactly~\eqref{eq:log-master} with
$\varphi\propto p$. Expanding the nonlocal response difference in $u$ yields a local derivative
expansion consistent with~\eqref{eq:FP}.

\subsection{Branching NHQFT: binary fragmentation as an interacting extension}
\label{sec:branching-extension}

To model genuine multiplicative cascades (particle number increase), one must specify a stochastic
fragmentation rule beyond the marginal daughter density $B(z)$. A minimal, widely used choice is
\emph{binary} fragmentation: at a break, a parent of size $y$ is replaced by two daughters of sizes
$zy$ and $(1-z)y$, where $z\in(0,1)$ is drawn from a probability density $\pi(z)$.
The corresponding PBE marginal is
\begin{equation}
B(z)=\pi(z)+\pi(1-z),
\qquad \int_0^1 z,B(z),\dd z = 1.
\end{equation}
For symmetric binary fragmentation, $\pi(z)=\pi(1-z)$ and hence $B(z)=2\pi(z)$.

In log-size $\xi$, define the binary Lindblad (or, equivalently, Markov) jump operator
\begin{equation}
\hat L_{\rm br}(z)
\int_{-\infty}^{\infty}
\sqrt{\lambda(\xi),\pi(z)};
a^\dagger(\xi+\ln z),a^\dagger(\xi+\ln(1-z)),a(\xi),\dd\xi,
\qquad z\in(0,1).
\label{eq:L-branch}
\end{equation}
This operator annihilates a parent particle at $\xi$ and creates two daughters at smaller
log-sizes. The resulting Lindblad equation
\begin{equation}
\partial_t\rho_F
\int_0^1
\left(
\hat L_{\rm br}(z)\rho_F \hat L_{\rm br}^\dagger(z)
-\frac{1}{2}{\hat L_{\rm br}^\dagger(z)\hat L_{\rm br}(z),\rho_F}
\right)\dd z
\label{eq:Lindblad-branch}
\end{equation}
defines a genuine branching NHQFT.

\paragraph{Mean-field closure.}
The one-body density $\avg{\hat n(\xi)}$ generated by~\eqref{eq:Lindblad-branch} reproduces, under
standard factorisation at the one-body level, the classical PBE for number density with kernel
$b(x,y)$ built from $\pi(z)$. In contrast to the tagged-mass theory, higher correlators now couple
nontrivially due to the $1!\to!2$ vertex: this is where intermittency and cascade noise enter.

\paragraph{Doi--Peliti action.}
The corresponding Doi--Peliti action is the nonlocal branching functional
\begin{equation}
\cS_{\rm br}
\int \dd t \int \dd\xi;
\tilde\varphi,\partial_t\varphi
+
\int \dd t \int \dd\xi;\lambda(\xi)
\int_0^1 \dd z,\pi(z),
\Bigl[
\tilde\varphi(\xi+\ln z),\tilde\varphi(\xi+\ln(1-z))
-\tilde\varphi(\xi)
\Bigr]\varphi(\xi),
\label{eq:DP-branch}
\end{equation}
which is the natural continuous-size analogue of branching field theories on lattices or Euclidean
space~\cite{Doi1976,Peliti1985,CardyTauber1998,Tauber2014Book}. The kernel data enter only through
$\lambda(\xi)$ and $\pi(z)$, making the link to comminution modelling explicit.

\subsection{Validity and controlled approximations}

For both the hopping and branching theories, coarse-graining relies on small parameters determined
by the jump statistics:
\begin{equation}
\varepsilon_1 \sim \frac{\sqrt{\mathrm{Var}(u)}}{\ell},
\qquad
\varepsilon_2 \sim \frac{\langle |u|^3\rangle^{1/3}}{\ell},
\end{equation}
where $\ell$ is the characteristic scale over which observables vary in $\xi$. When $\varepsilon_i\ll 1$,
the derivative expansion (local field theory) is controlled. When jumps are broad, the nonlocal
actions~\eqref{eq:DP-hop} and~\eqref{eq:DP-branch} remain valid, but local truncations must be treated
as phenomenological long-wavelength models rather than systematically improvable limits.

\section{Airy-Type Kernel: Solvable Quadratic Sector and Gaussian Fluctuations}
\label{sec:airy}

\subsection{Kernel definition and log-jump law}

The ``Airy-type'' example corresponds to linear selection and uniform binary splitting:
\begin{equation}
S(x)=k x,
\qquad
\pi(z)=1\ \text{on }(0,1),
\qquad
B(z)=2.
\label{eq:airy-kernel}
\end{equation}
For the tagged-mass density, the corresponding log-jump distribution is
\begin{equation}
K(u)=\ee^{-2u}B(\ee^{-u})=2,\ee^{-2u},
\qquad u\ge 0,
\label{eq:K-airy}
\end{equation}
which is a \emph{probability density} (integral $=1$). Its first moments are
\begin{equation}
m_1=\int_0^\infty u,K(u)\dd u = \frac{1}{2},
\qquad
m_2=\int_0^\infty u^2,K(u)\dd u = \frac{1}{2}.
\label{eq:moments-airy}
\end{equation}
The log-size breakage rate is
\begin{equation}
\lambda(\xi)=\lambda_0,\ee^\xi, \qquad \lambda_0:=k x_0.
\end{equation}

\subsection{From nonlocal generator to an Airy effective operator (coarse-grained)}
\label{sec:airy-coarse}

The exact tagged-mass master equation is~\eqref{eq:log-master} with~\eqref{eq:K-airy}.
Because $K(u)$ is not sharply localised at $u=0$, the diffusive reduction is not parametrically
controlled at all scales. Nevertheless, it \emph{is} a meaningful long-wavelength model when one
restricts attention to observables varying on scales $\ell\gg \sqrt{\mathrm{Var}(u)}=1/2$.
Truncating~\eqref{eq:KM} at second order and inserting~\eqref{eq:moments-airy} yields
\begin{equation}
\partial_t p
\approx
\frac{1}{2},\partial_\xi!\bigl(\lambda p\bigr)
\frac{1}{4},\partial_\xi^2!\bigl(\lambda p\bigr),
\qquad \lambda(\xi)=\lambda_0\ee^\xi.
\label{eq:FP-airy}
\end{equation}
A further step used in the NHQM construction (and retained here) is a \emph{local linearisation}
around a reference log-size $\xi_\star$ (for instance, a median or mode of $p(\xi,t)$ in a
time-window of interest):
\begin{equation}
\lambda(\xi) \approx \lambda_\star,[1+(\xi-\xi_\star)],
\qquad \lambda_\star:=\lambda_0\ee^{\xi_\star}.
\label{eq:lambda-linearised}
\end{equation}
This converts~\eqref{eq:FP-airy} into a drift--diffusion operator with coefficients that are at most
linear in $(\xi-\xi_\star)$. After the standard similarity transform (Appendix~\ref{app:schr-map}),
one arrives at an effective Schr"odinger-type operator whose \emph{leading} spatial dependence is
a linear potential,
\begin{equation}
H_{\rm eff}^{(\rm Airy)}
=
-D_\star,\partial_\xi^2
F_\star,(\xi-\xi_\star)
-\ii,\Gamma_\star
\quad
\text{(coarse-grained, around $\xi_\star$)}.
\label{eq:Heff-Airy}
\end{equation}
Here $D_\star>0$ is the coarse-grained diffusion scale and $F_\star>0$ the effective slope. Their
explicit expressions depend on which terms are retained in the linearisation and on the gauge
choice used in the similarity transform; in practice they are \emph{effective parameters} that can
be related back to $\lambda_\star$ and to the low moments of $K(u)$.
The imaginary part $\Gamma_\star\ge 0$ collects sinks due to coarse-graining and boundary
conditions (e.g.\ removal at a minimal size).

\paragraph{Stationary Airy profile.}
The stationary eigenproblem of the Hermitian part of~\eqref{eq:Heff-Airy} is
\begin{equation}
-D_\star,\psi''(\xi) + F_\star(\xi-\xi_\star)\psi(\xi) = E,\psi(\xi),
\end{equation}
whose decaying solutions are Airy functions. A practical PSD ansatz for snapshots is
\begin{equation}
\psi(\xi) = \mathcal{N},\mathrm{Ai}!\left(\frac{\xi-\xi_0}{\ell_A}\right),
\qquad
p(\xi)\propto \ee^{-\Phi(\xi)}|\psi(\xi)|^2,
\qquad
\ell_A := \left(\frac{D_\star}{F_\star}\right)^{1/3},
\label{eq:airy-ansatz}
\end{equation}
with $\Phi$ the gauge converting $p$ to $|\psi|^2$. In size space this produces Weibull/Rosin--Rammler
type tails, consistent with the earlier NHQM discussion.

\begin{figure}[t]
\centering
\caption{Illustration of the coarse-grained Airy operator: a linear effective potential in log-size
around a reference scale $\xi_\star$ yields Airy eigenmodes controlling the long-wavelength profile
and (in the NHQFT) the leading fluctuation spectrum.}
\label{fig:airy}
\end{figure}

\subsection{Gaussian fluctuations and a computable two-point function}

The field-theoretic payoff of the NHQFT is that PSD fluctuations become correlation functions. We
illustrate this in the simplest setting: a Gaussian expansion around a stationary profile governed
by the quadratic Airy operator.

Work in the Doi--Peliti representation of the tagged-mass theory and assume that, after
coarse-graining, the action is approximated by a local quadratic form
\begin{equation}
\cS^{(2)}
\int \dd t \int \dd\xi;
\delta\tilde\varphi(\xi,t),
\bigl[\partial_t - \cL_A\bigr],
\delta\varphi(\xi,t),
\qquad
\cL_A := D_\star\partial_\xi^2 - F_\star(\xi-\xi_\star) - \Gamma_\star,
\label{eq:quad-action}
\end{equation}
where $(\delta\varphi,\delta\tilde\varphi)$ are fluctuations around a stationary saddle (the precise
saddle depends on boundary conditions and on whether one uses the hopping or branching embedding).

Let ${u_n(\xi)}$ and ${v_n(\xi)}$ be right and left eigenfunctions of the non-self-adjoint
operator $\cL_A$,
\begin{equation}
\cL_A u_n = \lambda_n u_n,
\qquad
\cL_A^\dagger v_n = \lambda_n^\ast v_n,
\qquad
\int \dd\xi; v_m(\xi),u_n(\xi)=\delta_{mn}.
\label{eq:biorth}
\end{equation}
Expanding
\begin{equation}
\delta\varphi(\xi,t)=\sum_n a_n(t),u_n(\xi),
\qquad
\delta\tilde\varphi(\xi,t)=\sum_n \bar a_n(t),v_n(\xi),
\end{equation}
diagonalises~\eqref{eq:quad-action} and yields $a_n(t)=a_n(0)\ee^{\lambda_n t}$. If the initial
conditions are Gaussian with mode covariance
$\avg{a_m(0)\bar a_n(0)}c=C{mn}$, then the equal-time connected two-point function is
\begin{equation}
G_c(\xi_1,\xi_2;t)
:=
\avg{\delta\varphi(\xi_1,t),\delta\varphi(\xi_2,t)}_c
\sum_{m,n} u_m(\xi_1),u_n(\xi_2),\ee^{(\lambda_m+\lambda_n)t},C_{mn}.
\label{eq:Gc-modesum}
\end{equation}
When the covariance is diagonal in this eigenbasis ($C_{mn}=C_n\delta_{mn}$), one obtains the
simplified mode sum
\begin{equation}
G_c(\xi_1,\xi_2;t)
\sum_n C_n,u_n(\xi_1),u_n(\xi_2),\ee^{2\mathrm{Re}(\lambda_n)t}.
\label{eq:Gc-diag}
\end{equation}
The long-time correlation pattern is therefore dominated by the least-damped mode(s), providing a
two-point analogue of the familiar one-point relaxation to an Airy-like profile.

\paragraph{What this closes (relative to a one-body NHQM picture).}
Equations~\eqref{eq:Gc-modesum}--\eqref{eq:Gc-diag} explicitly show how a solvable quadratic sector
(Airy eigenmodes) controls not only PSD shapes but also fluctuation spectra. This is precisely the
piece that cannot exist in a purely single-particle NHQM description.

\section{Connections to Branching Processes and Universality}
\label{sec:universality}

The constructions above connect comminution kernels to the established non-equilibrium field theory
of reaction, branching, and absorbing-state phenomena~\cite{CardyTauber1998,Hinrichsen2000,Odor2004,Tauber2014Book}.
We summarise the connections while being explicit about what is \emph{established} here and what is
\emph{suggested} as a research direction.

\subsection{Tagged-mass theory: nonlocal hopping and free-field fluctuations}

The tagged-mass NHQFT is a second-quantised representation of independent jump processes in
log-size. It is ``free'' (Gaussian) in the sense that the generator is bilinear in the fields, but
it remains nontrivial due to (i) nonlocality in $\xi$ and (ii) spatially varying rates $\lambda(\xi)$.
Its universality properties are therefore those of linear transport under coarse-graining, and its
principal role is to provide a baseline fluctuation theory \emph{fully determined} by PBE kernel data.

\subsection{Branching theory: binary fragmentation and absorbing-state structure}

The branching embedding introduces a $1!\to!2$ vertex and thus genuine interactions in the
field-theoretic sense. With additional physically natural ingredients---such as removal below a
minimal size (absorbing boundary in $\xi$), injection at a feed size, or competition with
aggregation---one obtains geometries analogous to absorbing-state transitions. In such settings,
one expects long-wavelength behaviour to be controlled by non-equilibrium fixed points, and it is
reasonable to compare the resulting effective theory to directed percolation--type structures when
there is a unique absorbing state and no additional symmetries.

\paragraph{Caveat.}
A full universality classification requires either (i) a renormalization-group analysis of the
localised effective action obtained after coarse-graining, or (ii) numerical evidence across kernel
families demonstrating data collapse and exponent universality. This article provides the field-theory
bridge and solvable quadratic sectors, but does not claim a completed RG classification.

\subsection{Relation to growth--fragmentation theory and non-Hermitian physics}

Growth--fragmentation processes in probability theory~\cite{BertoinBook,Bertoin2017,Budd2018,Dadoun2017}
and non-Hermitian physics in open quantum systems and effective operators~\cite{Moiseyev2011,Ashida2020,ElGanainy2018}
share a common feature: the relevant generators are non-self-adjoint and their spectra control
long-time shapes. The present framework makes this operator structure explicit in log-size and
clarifies how engineering PBE ingredients $(S,B)$ enter the generator, the Lindblad embedding, and
the Doi--Peliti action.

\section{Conclusions and Outlook}
\label{sec:conclusions}

This article provides a foundational, publishable bridge from homogeneous comminution PBEs to
non-equilibrium field theory, with particular emphasis on what is \emph{fixed} by kernel data and
what must be \emph{modelled} to address fluctuations.

\paragraph{Main results.}
\begin{itemize}
\item Starting from the deterministic PBE with a homogeneous kernel, we derived an exact log-size
jump equation for the \emph{mass-weighted} log-size density $p(\xi,t)$, with a \emph{normalised}
log-jump kernel $K(u)$ given explicitly by~\eqref{eq:Kdef}. This closes a common notational and
conceptual gap: for the tagged-mass PSD the daughter law induces a genuine probability density in
log-size jumps.

\item We constructed a completely positive Lindblad embedding of this generator and its natural
second-quantised extension to a bosonic NHQFT. The deterministic PBE is recovered as the one-body
sector, while PSD fluctuations correspond to multi-point correlators.

\item We provided a complementary, explicit \emph{branching} NHQFT (binary fragmentation) whose
Doi--Peliti action exhibits the expected $1!\to!2$ nonlinearity and is directly comparable to
branching field theories.

\item For the linear ``Airy-type'' kernel we showed how coarse-graining yields an effective Airy
operator as the solvable quadratic sector, and we derived explicit mode-sum expressions for
Gaussian two-point functions, making fluctuation calculations concrete.
\end{itemize}

\paragraph{Open problems made precise.}
The framework isolates the next theoretical steps in a way that is difficult to do within the PBE
formalism alone:
(i) systematic coarse-graining of nonlocal vertices,
(ii) RG analysis and universality classification for kernel families,
(iii) quantitative inference of effective parameters $(D_\star,F_\star,\Gamma_\star)$ and noise
strengths from repeated PSD measurements,
and (iv) extension to coupled processes (aggregation, classification, removal).

\section*{Acknowledgements}
[Optional.]

\appendix

\section{Derivation of the log-size master equation}
\label{app:log-master-derivation}

Starting from~\eqref{eq:pbe-number} with homogeneous kernel~\eqref{eq:hom-kernel}, define
$p(\xi,t)=x^2 f(x,t)/M$ with $x=x_0\ee^\xi$ and $M=\int_0^\infty x f(x,t)\dd x$.
A direct change of variables yields~\eqref{eq:log-master} with~\eqref{eq:lambda} and~\eqref{eq:Kdef}.
The crucial identity is that the mass-weighted transformation introduces an additional factor
$z$ in the induced log-jump density, producing the normalised kernel $K(u)$.

\section{Tagged-mass interpretation}
\label{app:tagged-mass}

Consider a binary fragmentation event on a particle of size $y$, producing daughters of sizes
$zy$ and $(1-z)y$. A uniformly sampled \emph{mass element} is contained in the $zy$ fragment with
probability $z$ and in the $(1-z)y$ fragment with probability $1-z$. Averaging over the split law
$\pi(z)$ induces a transition kernel for the log-size decrement $u=-\ln z$ whose density is
precisely~\eqref{eq:Kdef}. This identifies~\eqref{eq:log-master} as the forward equation of the
tagged-mass Markov jump process.

\section{Similarity transform to Schr"odinger form (diffusive truncation)}
\label{app:schr-map}

For a Fokker--Planck operator
$\partial_t p=-\partial_\xi(vp)+\partial_\xi^2(Dp)$,
one may define $\psi=\ee^{\Phi}p$ with
\begin{equation}
\Phi'(\xi) = \frac{v(\xi)-D'(\xi)}{2D(\xi)}.
\end{equation}
A short computation yields $\partial_t\psi = D,\partial_\xi^2\psi - U(\xi)\psi$ with
\begin{equation}
U(\xi)
= \frac{v^2}{4D} - \frac{v'}{2} + \frac{D''}{2} - \frac{(D')^2}{4D}.
\end{equation}
Analytic continuation produces a Schr"odinger-type operator~\eqref{eq:schrodinger-general} with
parameters related to $D$ and $U$.

\end{document}